\documentclass[aps,prb,twocolumn,showpacs,superscriptaddress,amsmath,amssymb,citeautoscript]{revtex4-1}

\usepackage{graphicx}

\usepackage{color}

\usepackage{bm}
\renewcommand{\vec}[1]{\bm{#1}}

\begin{document}

\title{Fidelity at Berezinskii-Kosterlitz-Thouless quantum phase transitions}

\author{G. Sun}
\affiliation{Institut f\"ur Theoretische Physik, Leibniz Universit\"at Hannover, 30167~Hannover, Germany}

 \author{A. K. Kolezhuk}
\affiliation{Institute of High Technologies, Taras Shevchenko National University of Kiev,  03022 Kiev, Ukraine}
\affiliation{Institute of Magnetism, National Academy of Sciences and Ministry of Education,  03142 Kiev, Ukraine}

\author {T. Vekua}
\affiliation{Institut f\"ur Theoretische Physik, Leibniz Universit\"at Hannover, 30167~Hannover, Germany}

\begin{abstract}
We clarify the long-standing controversy concerning the behavior of
the ground state fidelity in the vicinity of a quantum phase
transition of the Berezinskii-Kosterlitz-Thouless type in
one-dimensional systems. Contrary to the prediction based on the
Gaussian approximation of the Luttinger liquid approach, it is shown
that the fidelity susceptibility does not diverge at the transition,
but has a cusp-like peak $\chi_c- \chi(\lambda)\sim
\sqrt{|\lambda_c-\lambda|} $, where $\lambda$ is a parameter driving
the transition, and $\chi_c$ is the peak value at the transition point $\lambda=\lambda_c$. Numerical claims of the
logarithmic divergence of fidelity susceptibility with the system size
(or temperature) are explained by logarithmic corrections due to
marginal operators, which is supported by numerical
calculations for large systems.
\end{abstract}
\date{\today}



\maketitle

\section{Introduction}

The ground state fidelity \cite{ZanardiPaunkovic06,VenutiZanardi07,You+07}, a concept stemming from
quantum information theory, 
is the overlap amplitude
$F(\lambda,\lambda+\delta\lambda)=|\langle \psi_0(\lambda)|
\psi_0(\lambda+\delta\lambda)\rangle|$ between two ground state wave functions of
the Hamiltonian $\hat{\mathcal{H}}(\lambda)=\hat{\mathcal{H}}_{0}+\lambda \hat{V}$
at different values of the coupling parameter $\lambda$.  
It is widely used as an unbiased
indicator of quantum phase transitions
\cite{You+07,VenutiZanardi07,Schwandt+09,Gu10rev}, especially in one-dimensional (1D)
systems where a very accurate numerical calculation of the ground state wave function is
possible thanks to the well-developed density matrix renormalization group
(DMRG) technique \cite{White,Uli}. Hereafter, we restrict the discussion to the
1D case. 

Fidelity vanishes
exponentially with the system size $L$. The fidelity susceptibility
per site (FS) 
\[
\chi_L=(1/L)\lim_{\delta\lambda\to0} \big[-2\ln
F(\lambda,\lambda+\delta\lambda)\big]/(\delta\lambda)^{2}
\]
is an intensive quantity expected to diverge in the thermodynamic
limit $L\to\infty$ at the phase transition point
$\lambda=\lambda_{c}$ due to nonanalyticity in the ground state. 
For finite $L$, this divergence translates into a presence of a peak in
$\chi_{L}(\lambda)$ at $\lambda=\lambda_{*}$, with $\lambda_{*}\to
\lambda_{c}$ and $\chi_{L}(\lambda_{*})\to\infty$ at
$L\to\infty$. 
Assuming a translational invariant system with the unique ground state,
perturbed by a local operator
$\hat{V}=\partial_{\lambda}\hat{\mathcal{H}}=\sum_{x}\hat{V}(x)$,
one obtains \cite{VenutiZanardi07} the
following connection 
\begin{equation} 
\label{fs-corr1} 
\chi_{L}=\int_{a}^{L}dx \int_{0}^{\infty} d\tau\, \tau  G(x,\tau),
\end{equation} 
between the FS and the reduced correlation function
$G(x,\tau)=\langle\!\langle
\hat{V}(x,\tau)\hat{V}(0,0)\rangle\!\rangle\equiv
\langle \hat{V}(x,\tau)\hat{V}(0,0) \rangle - \langle \hat{V}(0,0)
\rangle^{2},$
where the imaginary time evolution is defined by $\hat{V}(x,\tau)=e^{\tau
  \hat{\mathcal{H}}} \hat{V}(x) e^{-\tau
  \hat{\mathcal{H}}}$, averages are taken in the ground state
$|\psi_{0}(\lambda)\rangle$, and $a$ is the short-range (lattice) cutoff.
Expression (\ref{fs-corr1}) diverges at $L\to\infty$ as $\chi\propto
L^{1+2z-2\Delta_{V}}$, where $\Delta_{V}$ is the scaling dimension of
$\hat{V}(x)$ at the critical point and $z$ is the dynamic exponent, as long
as $\Delta_{V}<z+1/2$. At $\Delta_{V}=z+1/2$ there is only a logarithmic divergence
\cite{GuLin09}, and with the further increase of $\Delta_{V}$ the FS 
remains finite at the critical point. 

\section{Controversy}
\label{sec:controversy}

The above arguments~\cite{VenutiZanardi07} show that the FS must be
insensitive not only against marginal and irrelevant perturbations
($\Delta_{V}\geq z+1$), 
but even against relevant perturbations with  $z+1/2 <\Delta_{V}<z+1$. 
In the
case of the Berezinskii-Kosterlitz-Thouless (BKT) phase transition, $z=1$ and
$\Delta_{V}=2$, so it has been initially concluded \cite{You+07,VenutiZanardi07,Chen+08} that transitions of this type
cannot be detected by means of the finite-size scaling analysis of the FS. A
prominent example of the BKT transition is
the transition at the isotropic point $\lambda=1$ in the so-called XXZ
spin-$\frac{1}{2}$ chain defined by the Hamiltonian
\begin{equation} 
\label{ham-XXZ}
\hat{\mathcal{H}}_{XXZ}= \sum_{n} \Big\{ S^{x}_{n}S^{x}_{n+1}
+S^{y}_{n}S^{y}_{n+1} +\lambda S^{z}_{n}S^{z}_{n+1}\Big\},
 \end{equation}
where $S^{a}_{n}$ are spin-$\frac{1}{2}$ operators at site $n$. 

This conclusion has been apparently defeated by Yang \cite{MFYang07} and Fj{\ae}restad
\cite{Fjarestad08}. Their approach uses the
fact that the low-energy  effective theory of the model
(\ref{ham-XXZ}) (as well as of many
other gapless 1D systems), obtained by the Abelian bosonization
\cite{Giamarchi-book},  is the so-called Luttinger liquid (LL) described by the
Hamiltonian
\begin{equation} 
\label{ham-LL}
 \mathcal{H}_{\rm eff}=\frac{v}{2}\int dx \Big\{ K\Pi^{2}+ \frac{1}{K}(\partial_{x}\Phi)^{2}\Big\},
\end{equation}
where $\Phi$ is the compact bosonic field ($\Phi=\Phi+\sqrt{\pi}$), and $\Pi$ is its
conjugate momentum. The velocity $v=v(\lambda)$ and the LL parameter $K=K(\lambda)$ are
generally functions of the original coupling $\lambda$ that have to be
obtained as fixed points of the renormalization group (RG) flow equations. 
Alternatively, they can be extracted from the
knowledge of exact long-distance behavior of correlation functions. 
For the XXZ
model (\ref{ham-XXZ}), exact correlator asymptotics is known from the Bethe
ansatz, which yields $K=\pi/[2(\pi-\arccos \lambda)]$ and
$v=\pi\sqrt{1-\lambda^{2}}/(2\arccos\lambda)$ \cite{JohnsonKrinskyMcCoy73}.
Since the effective model (\ref{ham-LL}) is quadratic, one can explicitly calculate
the fidelity $F(K,K+\delta K)$ and obtain for the FS in the thermodynamic limit   
\begin{equation} 
\label{LL-yang} 
\chi|_{L=\infty} =  (\partial_{\lambda}K)^{2}/(8aK^{2}).
\end{equation}
 The dependence
$K(\lambda)$ is singular at $\lambda=1$, which leads to the divergence of the FS
$\chi\propto (1-\lambda)^{-1}$.

This direct calculation of overlaps might seem questionable since the connection
between the wave functions of the initial model and its fixed-point low-energy
theory is not so clear. Instead, one can use an alternative derivation due to Sirker \cite{Sirker10}
based on the relation (\ref{fs-corr1}). Indeed, using the effective Hamiltonian
(\ref{ham-LL}), the perturbation
$\hat{V}=\partial_{\lambda}\hat{\mathcal{H}}$ can be represented as
\begin{equation} 
\label{oper} 
\hat{V}=\frac{\partial_{\lambda}v}{v} \hat{\mathcal{H}}(\lambda) +
 \frac{v\partial_{\lambda}K}{2K}\int dx \, \hat{W}(x),
\end{equation}
where
\begin{equation}
\label{oper-W}
\hat{W}(x)= K\Pi^{2}- \frac{1}{K}(\partial_{x}\Phi)^{2}.
\end{equation}
The first term in (\ref{oper}) commutes with $\hat{\mathcal{H}}(\lambda)$
and thus gives no contribution into $\chi_{L}$ \cite{note1-temp}, but the second term contributes
the prefactor leading exactly to the form (\ref{LL-yang}). The scaling dimension of the
second term is $\Delta_{W}=2$, so the correlator in
(\ref{fs-corr1}) behaves as  
\begin{equation} 
\label{gxt}
 G(x,\tau)\sim A(\lambda)
v^{2}(x^{2}+v^{2}\tau^{2})^{-2},
\end{equation}
the integral in (\ref{fs-corr1}) is finite at
$L\to\infty$, and the divergence originates solely from the prefactor.

One can show that this singular behavior of
$K(\lambda)$ is a generic property of
any BKT transition. The general theory of a BKT transition is given by
 the Hamiltonian (\ref{ham-LL})
perturbed by the cosine term $-(u/a^{2})\cos(\sqrt{16\pi}\Phi)$, where $a$ is the
lattice cutoff and $u$ is the dimensionless coupling. The proximity to the
transition is controlled by the parameter $(\lambda_{c}-\lambda) \approx K-1/2-\pi u/2$. 
It follows from the RG equations
\cite{Kosterlitz74,PelissettoVicari13} that in the thermodynamic
limit the fixed point behavior is $K-1/2 \approx \sqrt{\pi u (\lambda_{c}-\lambda)}$, so the FS diverges. 
In a finite system, however, the RG flow should be stopped at the RG
scale $l\propto \ln(L/a)$, and the derivative of $K$ behaves as $(\partial K/\partial
\lambda)\approx (4\pi u/3)l$ at $l\to\infty$, so  according to (\ref{LL-yang})
the FS at the transition should scale as 
\begin{equation} 
\label{log2} 
\chi_{L}\propto u^{2}\ln^{2}(L/a). 
\end{equation}

Indeed, numerical results show that in the vicinity of the isotropic point the
FS exhibits a peak \cite{WangFengChen10} located at $\lambda>1$, which  moves very slowly towards $\lambda=1$ with
increasing $L$, and whose height grows with
$L$  much faster than corrections in powers of
$1/L$ could  explain \cite{note-wang}.  A similar behavior has been reported \cite{Sirker10} for the
finite-temperature FS, and the results were claimed to be consistent with
the $\ln^{2}(T_{0}/T)$ behavior, essentially of the same origin as
Eq.\ (\ref{log2}) (in the case of an infinite system at finite temperature,
the RG flow is stopped at the scale $l\sim\ln(T_{0}/T)$ where $T_{0}$ is some
nonuniversal energy scale of the order of the spin exchange energy).
Recently, for the XXZ chain a divergent FS similar to (\ref{LL-yang}) has been claimed
\cite{LangariRezakhani12} on the basis of the real-space quantum renormalization
group.
On the other hand, other authors did not see any divergence in the FS at
BKT-type transitions in the spin-$\frac{1}{2}$ XXZ chain \cite{Chen+08} and in bosonic
 Hubbard model \cite{Carrasquilla+13,Damski14}. Further, comparison of the FS calculated according
to (\ref{LL-yang}) with the numerical results shows \cite{Sirker10} that in
order to fit
the data one has
to assume the ultraviolet cutoff $a$ to be strongly $\lambda$-dependent even in
the vicinity of the free fermion point $\lambda=0$.
This controversy is aggravated by the fact that 
a logarithmic growth is difficult to distinguish numerically from the well known logarithmic
finite-size corrections at the BKT transitions.

\section{Resolution of Controversy}
\label{sec:resolution}

There is a subtle problem with the above derivation of a diverging
FS, immediately revealed by a closer look at
the representation (\ref{oper}): recalling
the original model (\ref{ham-XXZ}), we see that
$\hat{V}=\sum_{n}S^{z}_{n}S^{z}_{n+1}$, so  $\hat{V}(x)$ is a bounded
operator, while in Eq.\ (\ref{oper})  the bounded operator $\hat{W}(x)$  carries a
prefactor that diverges at $\lambda\to1$ independently of any ultraviolet
cutoff. This indicates that the divergence might be an artefact of the
effective representation (\ref{oper}), built on the Abelian
bosonization and becoming
inapplicable in the vicinity of the transition point. A similar problem
is known for the amplitudes of correlation functions calculated by
Lukyanov \cite{Lukyanov99}: the
amplitudes explode at $\lambda\to1$, meaning that the applicability
of the correspondent asymptotics is pushed to larger and larger
distances. The integral (\ref{fs-corr1}) is convergent at $L\to
\infty$, so divergences coming from such prefactors may be compensated
by terms neglected in Eq. (\ref{ham-LL}). 

Generally, one should not rely on divergences stemming from prefactors
of operators of the fixed point action (action where we neglect all
irrelevant terms in particular those that may and will lead to
cancellation of these spurious divergences in physical
quantities). However, infrared divergences due to the correlators of
the fixed point action are physical, since irrelevant terms, being
'harmless' at long distances, can not compensate them.

Assuming, from the boundedness of $\hat{V}(x)$, that the
amplitude $A(\lambda)$ in the correlator (\ref{gxt}) remains finite at $\lambda\to1$,
one returns to the initial conclusion that the FS at the transition stays finite as well.
Finite-size corrections to $\chi$ might be naively estimated in a
standard way by exploiting the conformal symmetry: substitution
$(x\pm iv\tau)\mapsto (L/\pi)\sinh\left[\frac{\pi}{L}(v\tau \pm ix)
  \right]$ in the correlator (\ref{gxt}), mapping infinite space-time onto
a stripe, yields
\begin{equation} 
\label{fss-away} 
\chi_{L}= \chi_{0}(\lambda)\big[ 1-\frac{\pi a}{L} + O(L^{-2})\big], \quad \chi_{0}(\lambda)\sim \frac{\pi A(\lambda)}{2a} .
\end{equation}
Although numerical results \cite{VenutiZanardi07,Chen+08,LangariRezakhani12} for
the model (\ref{ham-XXZ}) in the gapless phase $|\lambda|<1$ are indeed
consistent with (\ref{fss-away}), this type of finite-size scaling certainly
breaks down close to the observed peak in the FS
\cite{VenutiZanardi07,Chen+08,WangFengChen10}. The logarithmic modification of the correlator $G(x,0)\mapsto
\ln^{\alpha}(x)G(x,0)$, which can take place at the isotropic point $\lambda=1$, would affect the finite-size corrections 
in (\ref{fss-away}) only by changing them from $L^{-1}$ to
$L^{-1}\ln^{\alpha}(L)$, not solving the problem.

To obtain the correct finite-size scaling of $\chi$ close to the SU(2)-symmetric BKT
transition point $\lambda=1$, it is convenient to use non-Abelian bosonization
\cite{Affleck85}.

In non-Abelian approach, the low-energy theory of the model
(\ref{ham-XXZ}) is described by the Hamiltonian 
\begin{eqnarray}
\label{act-WZW}
\!\!\mathcal{H}= \!\mathcal{H}_{\rm WZW}-\!\! \int \! \!\frac{d x}{2\pi}
\Big\{  g_{\parallel} J_{0}\bar J_{0} +\frac{g_{\bot}}{2}(J_{+}\bar
J_{-}\!+\!J_{-}\bar J_{+}\!)\!  \Big\},
\end{eqnarray}
where 
 $ \mathcal{H}_{\rm WZW}$ corresponds to 
level $k=1$ SU(2) Wess-Zumino-Witten (WZW)
theory \cite{DiFrancesco},
 \begin{equation}
\label{ham-WZW}
\mathcal{H}_{\rm WZW}=\frac{2\pi v}{3} \int dx
\Big\{  :\vec{J}\cdot\vec{J}: +  :\vec{\bar{J}}\cdot\vec{\bar{J}}: \Big\},
\end{equation}
and the second term in (\ref{act-WZW}) describes marginal
current-current perturbation. 
Here  $\vec{J}$ and $\vec{\bar{J}}$ are the currents of left- and
right-movers, which are holomorphic functions of the complex
coordinates $z=x+ iv\tau$ and $z^{*}$, respectively, and $:\ldots:$ denotes normal ordering. Currents satisfy the Kac-Moody algebra
\begin{equation} 
\label{kac-moody}
[J^{a}(z),J^{b}(z')]=\frac{i}{2\pi}\delta_{ab}\delta'(z-z')+i\varepsilon_{abc}J^{c}(z)\delta(z-z'),
\end{equation}
and their two-point correlation functions evaluated with the
unperturbed WZW fixed point action (in an infinite-size system) are
given by
\begin{equation} 
\label{2-point} 
\langle J^{a}(z)J^{b}(0)\rangle=\frac{\delta_{ab}}{4\pi^{2}|z|^{2}},
\quad \langle J^{a}(z)\bar{J}^{b}(z')\rangle =0.
\end{equation}
For finite system size $L$ the first correlator in Eq. (\ref{2-point})
gets modified by the conformal substitution $z\mapsto
\frac{L}{\pi}\sin(\frac{\pi z}{L})$ and the second equation of
Eq. (\ref{2-point}) gets modified into $\langle
J^{a}(z)\bar{J}^{b}(z')\rangle \propto \frac{1}{L^{2}}$.  The
cartesian and spherical components of the currents are connected in a
standard way, $J_{\pm}=J^{x}\pm iJ^{y}$, $J_{0}=J^{z}$.

Running couplings
$g_{\parallel}$ and $g_{\bot}$ are governed by the following
BKT-type RG equations \cite{Zamolodchikov}:
\begin{equation} 
\label{rg-wzw} 
\frac{d g_{\parallel}}{dl}=-\frac{2g_{\bot}^{2}}{2-g_{\parallel}},\quad \frac{d g_{\bot}}{dl}=-\frac{2g_{\parallel}g_{\bot}}{2-g_{\parallel}},
\end{equation}
where $l$ is the RG scale. 

\begin{figure}[tb]
\includegraphics[width=0.36\textwidth]{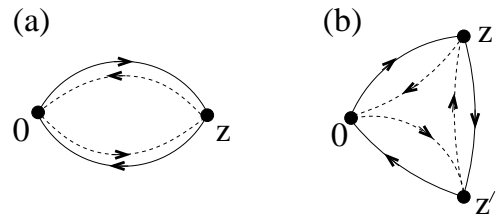}
\caption{  Diagrams for the current correlators contributing to
  $\chi_0$ (a) and $\chi_1$ (b), see Eqs.\ (\ref{chi0}), (\ref{chi1}), with continuous and dashed lines
  indicating left- and right-movers, respectively.
}
\label{fig:currentdiagrams}
\end{figure}

The fidelity-changing perturbation
 $\hat V= \sum_nS^z_nS^z_{n+1}$ can be  represented as 
\begin{eqnarray} 
\label{pert-split}
\hat
 V&=&\lambda^{-1}\Big\{\hat{\mathcal{H}}_{XXZ}-\sum_n(S^x_nS^x_{n+1}+S^y_nS^y_{n+1})\Big\} \nonumber\\
 &\mapsto& \lambda^{-1}\mathcal{H}- \int dx \hat{O}(x,\tau),
\end{eqnarray}
 where
\begin{eqnarray} 
\label{oper2}
\hat{O}(x,\tau) \sim  (J_{+}+\bar{J}_{+})(J_{-}+\bar{J}_{-})+\text{h.c.}
\end{eqnarray}
One can evaluate the correlator $G(x,\tau)=\langle\!\langle
\hat{O}(x,\tau) \hat{O}(0,0)\rangle\!\rangle$ 
in the perturbed WZW model (\ref{act-WZW}), and
calculate the fidelity susceptibility  $\chi_L=\int_{a}^{L} dx \int_{0}^{\infty} d\tau\, \tau G(x,\tau)$,
to the lowest order in the couplings $g_{\parallel}$, $g_{\bot}$.
Doing so, one obtains
\begin{equation}
\label{leadingcor-A}
\chi_L = \chi_{0}- g_{\parallel} \chi_{1}+O(g_{\parallel}^{2},g_{\bot}^{2}),
\end{equation}
where $\chi_{0}$, $\chi_{1}$ are finite positive constants discussed below. 

The
``unperturbed'' value $\chi_{0}$ of the FS (i.e., without taking into
account corrections due to marginal operators) is determined by the four-current
correlators (see the diagram shown in Fig. \ref{fig:currentdiagrams}(a)), that factorize into a
product of two-point functions such as $\langle\! \langle J_{+}(z) J_{-}(0) \rangle\!\rangle \langle\! \langle   \bar{J}_{-}(z) \bar{J}_{+}(0)\rangle\!\rangle$,
\begin{equation} 
\label{chi0} 
\chi_{0}\propto \int \frac{\tau \theta(\tau) \, d^2z}{|z|^4}>0.
\end{equation}
Above we have used the infinite-size current correlators
(\ref{2-point}). The constant $\chi_{0}$ here coincides with
$\chi_{0}(\lambda=1)$ in Eq.\ (\ref{fss-away}).  If one takes the
finite-size corrections into account by correcting current-current
correlators (\ref{2-point}) with the help of the conformal
substitution $z\mapsto \frac{L}{\pi}\sin(\frac{\pi z}{L})$, the value
$\chi_{0}$ gets modified according to
\begin{equation} 
\label{1overL} 
\chi_{0}\mapsto  \chi_{0}\Big(1-{\pi a/}{L}\Big)+O(1/L^2).
\end{equation}

The other constant 
\begin{equation} 
\label{chi1} 
\chi_{1}
\propto \int \frac{\tau \theta(\tau) \, d^2z \, d^{2}z'}{|z|^2 |z'|^{2}|z-z'|^{2}}
>0 
\end{equation}
is determined by six-current correlators that correspond to
three-point functions \cite{Affleck+89} of the marginal operator $\hat{O}$,
such as $ \langle\!\langle
J_{+}(z)   J_{0}(z')   J_{-}(0)  \rangle\!\rangle   \langle\!\langle  \bar{J}_{-}(z) \bar{J}_{0}(z') \bar J_{+}(0)
\rangle\!\rangle$ (see Fig. \ref{fig:currentdiagrams}(b)).

\subsection{Finite-size scaling}
\label{subsec:fss}

Combining Eqs.\ (\ref{leadingcor-A}), (\ref{1overL}), and the RG
equations (\ref{rg-wzw}) we can determine the
finite-size scaling of the FS in the vicinity of the BKT transition.

First we consider the scaling right at the transition,
i.e., at the isotropic point.
It is well known that finite-size corrections from marginal operators are
only suppressed logarithmically
\cite{Cardy86,Blote+86,Affleck86,Eggert+94} in the system size.
Indeed, at the SU(2) point $\lambda=1$ one has
$g_{\bot}=g_{\parallel}=g>0$, and for weak coupling the solution of RG
 equations simplifies to $g(l)\simeq 1/l$, where $l$ is the RG scale.
In the spirit of Ref.\ \onlinecite{Cardy86}, one can
stop the RG flow at the length scale $l_{L}\sim \ln(L/a)$, and
replace the running coupling $g(l)$ in Eq.\ (\ref{leadingcor-A})  by
its  ``RG-improved'' value $g(l_{L})\sim 1/\ln(L/a)$ taken at this scale, which yields
the following logarithmic finite-size scaling:
\begin{equation}
\label{logcor}
\chi_{L} \simeq \chi_{0} - \frac{\chi_{1}}{\ln(L/a)} + O\left[\frac{1}{\ln^2(L/a)}\right].
\end{equation}
We expect such scaling  of the FS to be valid at BKT transitions in
other 1D models as well, since any BKT transition point exhibits an enhanced
SU(2) symmetry.

Away from the SU(2) point inside the gapless region, log
corrections in (\ref{logcor}) get replaced by power laws. Indeed,
close to the SU(2) point the running coupling in the leading
correction to FS in Eq.\ (\ref{leadingcor-A}) can be RG-improved as
\cite{Lukyanov98} $ g_{\parallel}(l_{L})-g^{*}\propto (a/L)^{8K-4}$,
where the fixed point value of $g_{\parallel}$ is $g^{*}\equiv
g_{\parallel}(l=\infty)=2-K^{-1}$.  
Thus, for $K<\frac{5}{8}$
($\lambda>\cos\frac{\pi}{5}\approx 0.81$ in the XXZ chain) the leading
contribution to the 
finite size scaling of $\chi_{L}$ is given by
\begin{equation} 
\label{away-1} 
\chi_{L}-\chi_{\infty} \sim -(a/L)^{8K-4},\quad K<5/8,
\end{equation}
which transforms into log corrections in the SU(2) limit $K\to
\frac{1}{2}$.  For $K>\frac{5}{8}$, the above
correction becomes subleading with respect to the $1/L$ terms stemming from
(\ref{1overL}), and thus the leading finite-size correction is
\begin{equation} 
\label{away-2}
\chi_{L}-\chi_{\infty} \propto -(a/L),\quad K>5/8. 
\end{equation}

\subsection{Finite-temperature corrections}
\label{subsec:fts}

The above analysis is easily carried over to the low-temperature
behavior of FS. We adopt the definition of the finite-temperature FS
\cite{Schwandt+09} based on (\ref{fs-corr1}) with ground state
averages replaced by thermal ones, and the upper integration limit in
$\tau$ set to $1/(2T)$. Perturbative corrections to the FS will be again
determined by the formula (\ref{leadingcor-A}), with the
``RG-improved'' couplings
 taken at the RG cutoff scale $l_{T}\sim \ln(T_{0}/T)$, where
 $T_{0}\sim v$
 is some energy scale of the order of the exchange constant (set to
 unity in our Hamiltonian (\ref{ham-XXZ})).

Note that in this case the conformal
substitution $z\mapsto \frac{v}{\pi T}\sinh\big(\frac{\pi T z}{v}\big)$ 
in the correlator (\ref{gxt}) leads only to quadratic
finite-temperature corrections
of the type $\chi-\chi_{0}\propto -T^{2}$. However, there is another
contribution \cite{Sirker10} to the correlator stemming from the term 
in the perturbation that is proportional to the
Hamiltonian (the first term in Eq.\ (\ref{oper}), or
Eq.\ (\ref{pert-split})). This contribution is proportional to the
variance of the Hamiltonian, $\langle \mathcal{H}^{2}\rangle -\langle
\mathcal{H}\rangle^{2}$, and while it vanishes exactly in the ground
state (contrary to what is stated in Ref.\ \onlinecite{Sirker10}), 
the corresponding expression is nonzero in the case of thermal
averages. This  yields the following correction to the FS
\begin{equation} 
\label{linearT}
\chi \mapsto \chi +\frac{c_{v}}{8}\Big(\frac{\partial_{\lambda}v}{v}\Big)^{2},
\end{equation}
where 
 $c_{v}=(\pi T/3v)$ is the specific heat  of the model, so the correction is obviously positive and
linear in $T$. 
So, in the finite-temperature case Eq.\ (\ref{linearT}) can be viewed
as being formally similar to Eq.\ (\ref{1overL}), if one
substitutes $a/L$ by $T/T_{0}$ (up to the change of sign of the correction). 

In view of the above, the low-temperature behavior
of the FS can be deduced from Eqs.\ (\ref{logcor})-(\ref{away-2}) by
means of the substitution $(a/L)\mapsto (T/T_0)$.
Particularly, at the isotropic point
\begin{equation} 
\label{T-iso}
\left.\chi(T)\right|_{L=\infty}-\chi_0 \propto -1/\ln(T_0/T),\quad
T\to0.
\end{equation}
Away from the SU(2) point, the leading low temperature
correction to $\chi(T)|_{L=\infty}$ depends on $\lambda$: for $\lambda>\cos\frac{\pi}{5}$
the correction
is negative,
\begin{equation} 
\label{T-away-1} 
\left.\chi(T)\right|_{L=\infty}-\chi_0 \propto -(T/T_0)^{8K-4},
\end{equation}
while for
$\lambda<\cos\frac{\pi}{5}$ it becomes positive, 
\begin{equation} 
\label{T-away-2} 
\left.\chi(T)\right|_{L=\infty}-\chi_0 \propto T.
\end{equation}
Thus, the finite-temperature correction has to vanish at some value of
$\lambda$ close to $\cos\frac{\pi}{5}$.
This explains crossing of FS
curves corresponding to different low temperatures at $\lambda \simeq
\cos\frac{\pi}{5}$ in Fig.\ 4(a) of Ref.\ \onlinecite{Sirker10}.

\subsection{The FS behavior in the thermodynamic limit}
\label{subsec:shape}

Finally, we would like to discuss the shape of the FS curve 
 $\chi_{L}(\lambda)$ in the vicinity of the
BKT transition at $\lambda=\lambda_{c}$, in the thermodynamic limit
$L\to\infty$. Right at the transition, the FS is finite, $\chi_{L\to\infty}(\lambda_{c})=\chi_{0}$.
Off the transition point inside the gapped region, we can again use
Eq.\ (\ref{leadingcor-A}) to calculate the leading correction to the FS.
In the gapped region, there is a finite correlation length
$\xi\sim
e^{B|\lambda-\lambda_c|^{-1/2}}$, where $B$ is some positive numerical
factor, so for $L\gg \xi$ the RG flow gets stopped at the scale $l_{\xi}\sim \ln(\xi/a)$,
and, similar to Eq.\ (\ref{logcor}), we obtain the leading
contribution as
\begin{equation}
\chi_{0}- \chi_{\rm gapped} (\lambda) \propto \frac{1}{l_{\xi}}
 \propto  |\lambda -\lambda_c|^{1/2} .
\end{equation}
We note that in the gapped region there is an additional factor
$\exp\{-(x^2+v^2\tau^2)^{1/2}/\xi\}$ in the correlator (\ref{gxt}), but
it only introduces a negligible (very smooth) modification of the FS in the
vicinity of BKT point, since the correlation length diverges exponentially
at the transition.

On the other side of the BKT transition, inside the gapless region,  
as we have already seen in Sec.\ \ref{subsec:fss}, the coupling $g_{\parallel}$ flows to the
finite fixed point value $g^{*} =2-1/K$, and
finite-size corrections vanish according to the power law
$ g_{\parallel}(l_{L})-g^{*}\propto  (a/L)^{8K-4}$.
Hence, in the thermodynamic limit  $L\to\infty$ one has
\begin{eqnarray}
\chi_{0}- \chi_{\rm gapless} (\lambda) &\propto&  (2-1/K) \propto  |\lambda -\lambda_c|^{1/2}.
\end{eqnarray}

Thus, in the thermodynamic limit, in the vicinity of the BKT
transition, the FS exhibits a peak with the square-root cusp:
\begin{equation}
\chi_{L=\infty}(\lambda) =\chi_{0}-\alpha_{\pm}
\sqrt{|\lambda -\lambda_c| }, \quad \lambda\to \lambda_{c}\pm0,
\end{equation}
where $\alpha_{\pm}$ are some positive factors.

It should be mentioned that the cusp must be rather
difficult to observe numerically: finite-size corrections tend to
smoothen the cusp, so one has to study systems up to a sufficiently
large size $L$. Another remark concerns the problem of defining the FS in the thermodynamic limit if the
ground state is degenerate, as is the case, e.g., for the
doubly degenerate N\'eel ground state of the XXZ chain in
the gapped region. In the next section we, in addition to the XXZ
chain, will consider two other models with the BKT transition that
have unique gapped ground states.



\begin{figure}[tb]
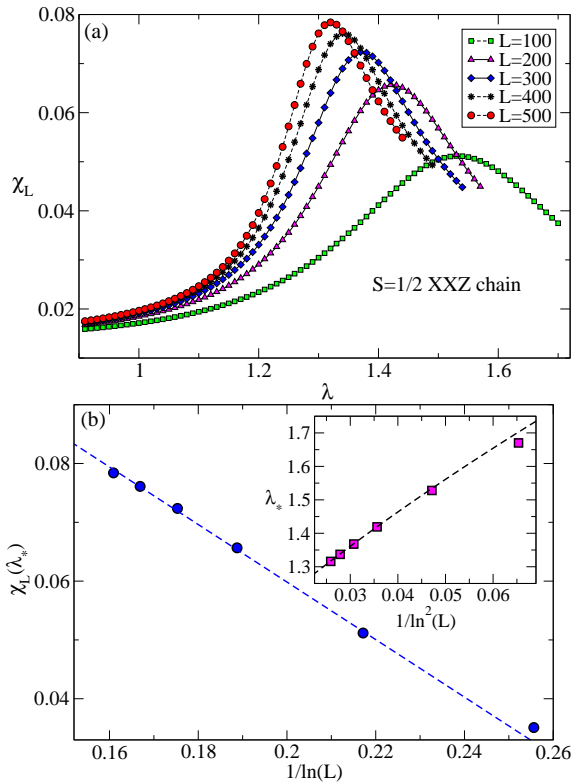

\includegraphics[width=0.42\textwidth]{fs-lambda-xxz}

\includegraphics[width=0.42\textwidth]{xxz-scaled}
\caption{ (Color online). (a) The fidelity susceptibility for spin-$\frac{1}{2}$
  XXZ chain (\ref{ham-XXZ}) of up to $L=500$ spins, as a function of the
  anisotropy $\lambda$; symbols denote DMRG results, and lines are guide to the
  eye.  (b) The finite-size scaling of the peak position $\lambda_{*}$ and
  amplitude $\chi_{L}(\lambda_{*})$; lines are results of fits to
  (\ref{logcor}) and (\ref{position-fit}), see text for details.
}
\label{fig:xxz-fits}
\end{figure}


\begin{figure}[tb]
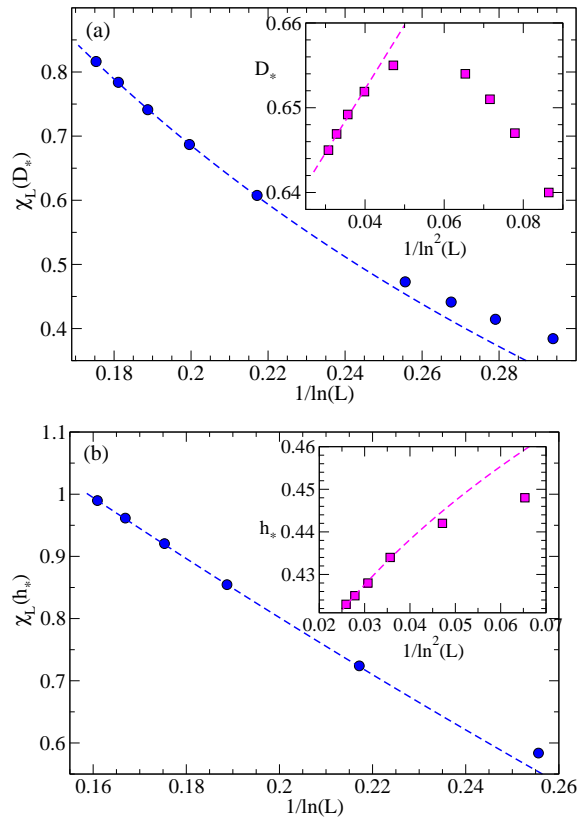

\includegraphics[width=0.42\textwidth]{s1bilbiq-scaled}

\vspace*{2mm}

\includegraphics[width=0.42\textwidth]{stagg-scaled}
\caption{ (Color online). (a) The finite-size scaling of the peak position
  $D_{*}$ and its value
  $\chi_{L}(D_{*})$, for the FS with respect to the  single-ion anisotropy $D$
  in the spin-$1$
   chain model defined by the Hamiltonian (\ref{ham-bilbiq}) with
   $\theta=-0.85\pi$;   (b) the same for the peak position $h_{*}$ and value
  $\chi_{L}(h_{*})$ of the FS with respect to the staggered field $h$ in the
   spin-$\frac{1}{2}$ anisotropic chain described by the model (\ref{ham-stagg}) at $\lambda=-0.9$.
Symbols show numerical (DMRG) results, and lines are fits to scaling
laws (\ref{logcor}), (\ref{position-fit}).
}
\label{fig:other-fits}
\end{figure}

\section{Numerical results} 
\label{sec:numerics}

To support our conclusions, we have performed DMRG calculations
\cite{White,Uli} (in matrix product formulation \cite{Verstraete})
of the FS for the model (\ref{ham-XXZ}), for
large open chains of up to $L=500$ sites \cite{note-dmrg}.
The resulting FS as a function of
$\lambda$ is shown in Fig.\ \ref{fig:xxz-fits}(a): as observed in previous
studies for small systems, there is a peak with the height slowly
growing and the position slowly
converging  with the increase of $L$. We start by fitting the finite-size dependence of the peak position $\lambda_{*}$ with the help of
the ansatz
\begin{equation} 
\label{position-fit} 
\lambda_{*}\simeq\lambda_{0}+\lambda_{1}/\ln^{2}(L/a)+\cdots,
\end{equation}  
which can be extracted by using standard scaling arguments\cite{SchultkaManousakis94} on the \emph{gapped}
side of the BKT transition: since the infinite-system correlation length in the
vicinity of the transition behaves as
$\xi\propto a e^{-B/\sqrt{\lambda-\lambda_{c}}}$, Eq.\ (\ref{position-fit}) is
obtained by postulating that $\xi\sim L$
at $\lambda\sim \lambda_{*}$. Further, when fitting the peak position
$\lambda_{*}$ according to (\ref{position-fit}), we fix $\lambda_{0}=1$, which allows us
to extract the  cutoff $a$. 
 Subsequently, we use the extracted value of the cutoff when fitting the
peak value $\chi_{L}(\lambda_{*})$ of the FS according to our result
(\ref{logcor}). The results of those fits, shown in
Fig.\ \ref{fig:xxz-fits},  demonstrate good agreement with the theory.

The above picture of non-diverging FS with strong
logarithmic corrections due to marginal operators should be a generic feature of any BKT transition. To
demonstrate that, we present here results of numerical studies for two more
models containing such transitions. The first model is the anisotropic spin-1 chain  defined by the Hamiltonian
\begin{equation}  
\label{ham-bilbiq} 
\hat{\mathcal{H}}=\sum_{n}\{  J_1^{\theta} (\vec{S}_{n}\cdot\vec{S}_{n+1}) 
+J_2^{\theta} \big(\vec{S}_{n}\cdot\vec{S}_{n+1}\big)^{2} +D\big(S_{n}^{z}\big)^{2}\}, 
 \end{equation} 
where $\vec{S}_{n}$ are spin-1 operators at site $n$,
$J_1^{\theta}=\cos{\theta}$ and $J_2^{\theta}=\sin{\theta}$ are
exchange constants, and $D>0$ is the
single-ion anisotropy. For $\theta\in [-\frac{3\pi}{2},-\frac{3\pi}{4}]$, with
the increase of $D$ this
model exhibits a BKT transition from the gapless ferromagnetic XY phase to the
gapped large-$D$ phase, recently studied \cite{Rodriguez+11} in the context of spinor
bosons. Numerically, the FS as function of $D$
exhibits a slowly growing peak \cite{Rodriguez+11}, quite similar to the
picture shown in Fig.\ \ref{fig:xxz-fits}(a). Here we extend the result of
Ref.\ \onlinecite{Rodriguez+11} to much larger systems \cite{note-selfpawn} and show that the finite-size behavior of the peak height and
position is consistent with the scaling formulas (\ref{logcor}) and
(\ref{position-fit}), see
Fig.\ \ref{fig:other-fits}(a).

One more model corresponds to the spin-$\frac{1}{2}$ ferromagnetic anisotropic chain defined by the
Hamiltonian (\ref{ham-XXZ}) with $-1< \lambda <0$, perturbed  by staggered field $h$:
\begin{equation} 
\label{ham-stagg} 
\hat{\mathcal{H}}=\hat{\mathcal{H}}_{XXZ}+h\sum_{n}(-1)^{n}S_{n}^{z}.
\end{equation}
For $-1< \lambda<-1/\sqrt{2}$, this model exhibits a BKT-type phase transition \cite{stag-xxz} to a
gapped antiferromagnetic phase at a finite value of the field $h$. Our
DMRG calculations for the FS as function of $h$ show the same typical behavior
of a slowly growing and poorly converging peak, and the 
finite-size scaling results 
presented in Fig.\ \ref{fig:other-fits}(b) show that the numerical
data is again consistent with the scaling laws (\ref{logcor}), (\ref{position-fit}).  

\section{Summary} 

We have shown that FS does not diverge at
Berezinskii-Kosterlitz-Thouless-type quantum phase transitions in one spatial
dimension. Instead, it merely exhibits a finite-amplitude peak in the vicinity
of the transition, with
logarithmic finite-size scaling
corrections of the form (\ref{logcor}), (\ref{position-fit}) which are too easy
to confuse numerically with a logarithmic growth of the peak. The same is true
for the finite-temperature FS, which instead of the claimed  \cite{Sirker10}
divergence $\chi\propto\ln^{2}(T_{0}/T)$ at $T\to 0$ should contain log
corrections of the form $\chi\simeq \chi_{0} -\chi_{1}/\ln(T_{0}/T)$.

The would-be divergence of the FS, originally proposed \cite{MFYang07}
on the basis of mapping to Luttinger liquid (LL), and in the meantime
enjoying the status of an established result
\cite{Venuti+08,Sirker10,LangariRezakhani12,Rey}, is an artefact due
to the catastrophic shrinking of the applicability range for using the
LL Gaussian effective description (\ref{ham-LL}) when calculating the
FS of the original microscopic model as one approaches the BKT
transition. To identify the correct scaling of the FS with respect to
some perturbation in a specific model at the BKT transition, it is
crucial to properly identify perturbation in effective description
that includes irrelevant corrections to the LL Guassian Hamiltonian
beyond the mere renormalization of LL parameters.

From a more general viewpoint, the main message of this work is a
warning about the naive use of the operator prefactors ("amplitudes")
of the effective action, obtained in the Abelian bosonization: one
should not trust divergences stemming from such amplitudes, since they
will be ``healed'' by the effects of irrelevant operators not taken
into account in the fixed point action.

On the practical side, our results indicate that using the FS as a
tool to detect BKT transitions (and especially to extract
thermodynamic critical values of microscopic parameters) is extremely
inconvenient, since the uncertainties of logarithmic fits remain too
strong, even if one goes to the largest numerically tractable system
sizes. Other detection methods suggested in quantum information
theory, e.g., looking at discontinuities of fidelity
\cite{WangChenLiZhou12}, or using bipartite fluctuations \cite{Rachel}
might be a better alternative.

\acknowledgments
 T.V. acknowledges motivating discussions with F.\ W.\ Diehl. This work has been supported by QUEST (Center for
Quantum Engineering and Space-Time Research) and DFG Research Training Group
(Graduiertenkolleg) 1729.



\begin{thebibliography}{10}
\bibitem{note1-temp}  This term does contribute to the finite-temperature
  FS, see Ref.\ \onlinecite{Sirker10}, but its contribution vanishes exactly for the ground state
  FS even in finite-size systems, contrary to the statement in Ref.\ \onlinecite{Sirker10}.
\bibitem{note-wang} The authors of Ref. \cite{WangFengChen10} attempted a
  power-law fitting of the FS which resulted in an anomalously low power
  $\chi\propto L^{0.025}$ usually indicating the presence of log corrections.
\bibitem{note-dmrg} In our calculations of the FS, we have chosen the
  step $\delta \lambda=10^{-3}$ and checked that difference between
  results with $\delta \lambda=10^{-4}$ is negligible. We have taken
  care that for any system size $L$, a convergence in the matrix
  dimension $m$ had been reached; about $m \lesssim 1000$ were
  typically sufficient to achieve good accuracy.
\bibitem{note-selfpawn} Particularly, in Ref.\ \onlinecite{Rodriguez+11} the
peak position exhibited an excellent $1/L^{2}$ scaling for system sizes $L<50$,
while our data show that the correct logarithmic behavior (\ref{position-fit}) sets in only for
$L\gtrsim 150$, see the inset of Fig. \ref{fig:other-fits}(a).
\end{thebibliography}
\end{document}